\documentclass{WileyMSP-template}

\usepackage{bbold}
\usepackage{mathptmx}
\usepackage{subfig}
\usepackage{psfrag,graphicx}
\usepackage{dcolumn}
\usepackage{amsmath}
\usepackage{amssymb}
\usepackage{amsfonts}
\usepackage{bm}
\usepackage{color}
\usepackage{latexsym}
\usepackage{epstopdf}
\usepackage{color}
\usepackage[english]{babel}
\usepackage{latexsym}
\usepackage{bm}
\usepackage{appendix}
\usepackage{rotating}
\usepackage{aeguill}
\usepackage{ulem}
\usepackage[justification=justified]{caption}

\usepackage[colorlinks=true,citecolor=blue,linkcolor=blue]{hyperref}

\newcommand{\ket}[1]{|{#1}\rangle}                       
\newcommand{\bra}[1]{\langle {#1}|}                      

\newcommand{\ketbra}[2]{\left\vert#1\right\rangle\left\langle#2\right\vert}
\newcommand{\projector}[1]{\ketbra{#1}{#1}}
\newcommand{\Ignore}[1]{ }

\begin{document}


\title{Axion-induced oscillations in Josephson qubits}

\maketitle


\author{Roberto Grimaudo*,}
\author{Claudio Guarcello,}
\author{Giovanni Filatrella,}
\author{Bernardo Spagnolo,}
\author{Davide Valenti}


\begin{affiliations}

Dr. R. Grimaudo, Prof. B. Spagnolo, Prof. D. Valenti\\
Dipartimento di Fisica e Chimica ``Emilio Segr\`{e}",
Universit\`{a} degli Studi di Palermo, viale delle Scienze, Ed. 18, I-90128, Palermo, Italy\\
Email: roberto.grimaudo01@unipa.it

Dr. Claudio Guarcello\\
Dipartimento di Fisica ``E.R. Caianiello'', Universit\`a di Salerno, Via Giovanni Paolo II, 132, I-84084 Fisciano (SA), Italy;
INFN, Sezione di Napoli Gruppo Collegato di Salerno, Complesso Universitario di Monte S. Angelo, I-80126 Napoli, Italy\\

Prof. G. Filatrella\\
Dep. of Sciences and Technologies and Salerno unit of INFN, University of Sannio, Via Port'Arsa 11, Benevento I-82100, Italy\\

Prof. B. Spagnolo\\
Lobachevskii University of Nizhnii Novgorod, 23 Gagarin Ave. Nizhnii Novgorod 603950 Russia

\end{affiliations}



\begin{abstract}

An effective quantum description of the axion is proposed as a two-level dynamic system together with a spin-spin Hamiltonian for a coupled axion-Josephson qubit system.
The interaction between axion and Josephson Junction can be responsible for a resonance effect, when the axion and the  Josephson frequencies match.
Thus, experimentally detectable oscillations induced in the Josephson qubit by its interaction with the axion are clearly highlighted.
This phenomenon, causing a periodic magnetization reversal in the junction, can be exploited for the axion detection in the quantum limit of low noise.

\end{abstract}

\section{Introduction}

Several observations, ranging from the anisotropies of the cosmic microwave background radiation to the dynamics of galaxy clusters, individual galaxies, and gravitational lensing, strongly suggest the existence of dark matter~\cite{Tanabashi18}, with an energy density fraction in the Universe falling in the range $26-27\%$~\cite{Tanabashi18,Banik13}.

The nature of dark matter is one of the outstanding problems in science today. The idea of the axion field and its fundamental excitations, the axions indeed, was introduced by Peccei and Quinn to propose a solution to the strong charge-parity problem in quantum chromodynamics (QCD), that is the absence of charge-parity violation in the strong interaction between quarks~\cite{Pec77}.
Indeed, among different ultralight dark matter models~\cite{Marsh16}, the leading candidate as elementary particle constituting the dark matter is the `QCD axion'.
The axion field is characterized by a potential $V(\theta)$ which stems from the strong interaction and has CP conserving minima~\cite{Weinberg78,Wilczek78}.

Another key feature of such dark matter particles is that they should generate an additional term in the laws of electrodynamics, causing to transform the electric and magnetic fields into each other~\cite{Wilczek87}.
Consequently, such `corrected' laws of electrodynamics should be observable in condensed-matter systems~\cite{Wilczek87, Nenno20}.

The relevance and the growing interest toward the axion search, both theoretical and experimental, is proved by the remarkable increase of works in the last 
years~\cite{ Nenno20, Gomez20, Arvanitaki20, Battye20, Braine20, Buschmann20, Co20, Backes21, Berlin21, Chaudhuri21, Nagano21, Wang21,Xiao21}. A wide class of experiments 
has been proposed in order to probe the presence of axions~\cite{Alesini21, Salemi21, Reynolds20, Dessert20, Kohri17, Moroi18, Caputo19, Gramolin21, Khlopov18}.
These experiments are based on the axion-photon conversion (Primakoff effect) under an external magnetic field, due to a coupling between axions and photons
~\cite{Sik83, Raffelt88, Murayama}.
Very recently, the interaction between a Josephson qubit and the axion-induced photons in a resonant cavity has been 
analysed~\cite{Dixit21}.

Josephson junctions (JJs), of paramount importance in many fields \cite{Dev84,Guarcello15,Lee20,Wal21}, play a foremost role in the search of possible protocols and schemes for the axion detection. 
Among the methods of detection proposed, Josephson-based experiments, grounded on a direct interaction between axions and JJs~\cite{Bec13,Bec17}, may help to understand unclear `events'~\cite{Hof04, Bae08, Gol04}.
However, in spite of the intense efforts, no direct observations have been registered in terrestrial experiments until now.

In the present work we change the paradigm by proposing a quantum description of the axion as an effective two-level system (TLS) that is appropriated in an environment characterized by negligible noise.
The model is based on the analogy with the Josephson systems.
The equations of motion governing the two systems, axion and JJ, are, indeed, formally identical, as well as the related anharmonic potentials that can be defined~\cite{Murayama, Alexandre}.
Since, under specific conditions, the JJs actually behave as TLSs~\cite{Blais21, Krantz}, it is possible to assume that, if similar conditions are satisfied by the axion field, the same quantum description can be suitable for the axions.

It is worth pointing out that a quantum description of the axion field has been proposed years ago and it has been proved that the axion potential is flattened if only gravitational self-interactions of the axion field are considered.
Nevertheless, interactions of axions with other particles prevent such a flattening effect~\cite{Alexandre}.

Here, we take into account the model describing a direct (i.e. not mediated by a cavity) interaction between axions and JJs~\cite{Gri22}.
This model is formally identical to that of two capacitively coupled JJs.
Taking into account the quantized version of both the axion and the JJ, we consider a model of two interacting TLSs in analogy with the quantized two-JJ-system model.
In this framework, then, the Josephson qubit directly interacts with two-level axion systems, rather than the quantized boson field induced by the axion-photon conversion, as in~\cite{Dixit21}.

The main signature of the axion presence is the occurrence of axion-induced oscillations in the Josephson qubit, due to the resonant matching between the axion and the JJ frequencies.
This effect might be appropriately observed and exploited to reveal the axions through their direct coupling with Josephson systems, provided that the quantum regime during the measure is preserved.

\section{Josephson Junctions and Axions}

\subsection{Josephson junctions}

The dynamics of the Josephson phase can be described within the resistively and capacitively shunted junction (RCSJ) model~\cite{McC68,Bar82,Gua15,Gua17,Gua19,Gua20},
\begin{equation}
\frac{d^2 \varphi}{dt^2}+ {1 \over RC} \frac{d \varphi}{dt} + {2 \pi \over \Phi_0} {I_c \over C} \sin \left[ \varphi\left( t \right) \right] = 0.
\label{RCSJnormOc}
\end{equation}
Here, $I_c$ is the maximum Josephson current that can flow through the device, $i_b= I_b / I_c$ is the normalized external bias current and $R$ and $C$ are the 
normal-state resistance and capacitance of the JJ, respectively. $e$ is the electron charge and $\Phi_0 = h/2e$ is the superconducting magnetic flux quantum.

The quantum Hamiltonian of a JJ can be written as
\begin{equation} \label{Quantum Ham JJ}
H = 4 E_c \hat{n}^2 - E_J \cos(\hat{\varphi}),
\end{equation}
where $\hat{n}$ and $\hat{\varphi}$ represent the charge and phase non-commuting operators, respectively, while $E_c = e^2/2C$ and $E_J = I_c \Phi_0 / 2 \pi$.
If $E_J \gg E_c$ ($E_J \ll E_c$), the phase (charge) is a `good' quantum number, that is the phase (charge) has a precise value while the charge (phase) is completely 
undetermined. Under the conditions $E_J \gg E_c$ and $E_J \gg k_B T$, the JJ effectively behaves like a TLS (qubit) and thus it can be formally described in terms of 
spin-1/2 dynamical variables~\cite{Blais21, Krantz}. This is possible thanks to the characteristic anharmonicity of the Josephson systems stemming from the sinusoidal 
term in Eq.~\eqref{Quantum Ham JJ}. The Hamiltonian of the effective qubit system, called phase qubit, reads 
\begin{equation}
H_J = \hslash \omega_J \hat{\sigma}_J^z,
\end{equation}
where $\hat{\sigma}_J^z$ being the well known spin-1/2 Pauli operator and $\omega_J = {(\sqrt{8E_J E_c}-E_c)/\hslash} = \omega_p - E_c/\hslash$ the effective qubit frequency.
The latter comes from the Lamb-shift-induced correction $E_c/\hslash$ to the plasma frequency $\omega_p=\sqrt{8E_JE_c}/\hslash$.

If $E_J \ll E_c$, the Josephson dynamics is analogous to that of a TLS as well, and a charge-qubit is generated.
However, in this regime, the effective qubit system becomes highly sensitive to the charge noise, which has proved more challenging to reduce than the flux noise.

\subsection{Axions}
The axion field is $a = f_a \theta$, where $f_a$ and $\theta$ are the axion coupling constant and the misalignment angle, respectively.
Within the Robertson-Walker metric, which is appropriate to describe the early universe, the equation of motion of $\theta(t) = a(t)/f_a$ 
reads~\cite{Co20}
\begin{equation}
\frac{d^2 \theta (t)}{dt^2}+ 3 H \frac{d \theta (t)}{dt} + \frac{m_a^2c^4}{\hslash^2} \sin \left [ \theta \left ( t \right ) \right ] = 0,
\label{AxionEq}
\end{equation}
where spatial gradients are neglected~\cite{Gri22}. Here, $H \approx 2 \times 10^{-18} ~ s^{-1}$ is the Hubble parameter and $m_a$ denotes the axion mass. The forcing term $\sin(\theta)$ is due to quantum chromodynamics instanton effects.
The effective axion potential is defined as
\begin{equation}
   V(\theta)={m_a^2c^4 \over \hslash^2}(1-\cos{\theta}).
\end{equation}
The misalignment mechanism explains the emergence of the axion mass as due to the initial non-equilibrium position [$\theta(0) \neq 0$], which induces oscillations of the 
axions around the potential minimum~\cite{Co20}. The axion, therefore, can be effectively described as a particle in an anharmonic potential well, whose anharmonicity 
stems from the $\cos(\theta)$ term, implying that for small oscillations [$V(\theta) \propto m_a\theta^2/2$] the axion dynamics is analogous to that of a harmonic oscillator.

This effective dynamic description is analogous to that of a JJ, since the analytical form of the potential is the same. Further, the similarity between the axion field and the 
JJ equation is evident: the axion dynamics is analogous to that of an RCSJ with no externally applied bias current.
Moreover, it is worth to remark that, besides the formal mathematical analogy between the two systems, the normalized parameters characterizing the two equations are quite similar as their order of magnitude is concerned~\cite{Gri22}.

As for JJs, the axion degrees of freedom can be quantized.
In this framework, therefore, it can be argued that the anharmonicity of the \textit{cosine} term induces a nonconstant spacing of the quantized  energy levels in the potential well.

In the light of the previous observation, by comparing the axion equation with that of the JJ and by establishing a correspondence between the analogous terms, it is possible 
to verify that the axion can be regarded as a JJ characterized by a plasma frequency $ m_ac^2/\hslash \equiv \omega_a \approx 100$ GHz, which is very close to the typical values of the Josephson plasma frequencies. This frequency, both for the axion and the JJ, corresponds to the frequency of small oscillations at the bottom of a well of the \textit{cosine} potential.
Based on this analogy, in the low-temperature regime $\hslash \omega_a \gg k_BT$, we can suppose that also the axion can be approximately considered a TLS, that is a system whose dynamics is basically restricted to the two lowest quantized levels, with the following effective Hamiltonian
\begin{equation}
H_a = \hslash \omega_a \hat{\sigma}_a^z,
\end{equation}
where $\hat{\sigma}_a^z$ is the Pauli operator associated to the axion.

\section{Axion-JJ System}

Several paradigms, aimed both at the search and at the detection of axionic dark matter in the halo of our galaxy, are based on the coupling between axion and photon~\cite{Murayama,Gri22,Bradley03}.
In analogy to what happens in resonant cavities, the axion-JJ coupling is supposed to be responsible for the decay of the axion in photons. 
The axion-JJ interaction can be formally written as
\begin{subequations}\label{Orig Diff Eqs Syst}
\begin{align}
\ddot{\varphi} + a_1 \dot{\varphi} + b_1 \sin(\varphi) &= \gamma (\ddot{\theta} - \ddot{\varphi}) \label{Orig Diff Eqs Syst a},\\
\ddot{\theta} + a_2 \dot{\theta} + b_2 \sin(\theta) &= \gamma (\ddot{\varphi} - \ddot{\theta}), \label{Orig Diff Eqs Syst b}
\end{align}
\end{subequations}
where $(a_1, a_2)$ and $(b_1, b_2)$ are the dissipation and frequency parameters, respectively~\cite{Gri22}; $\gamma$ is the coupling constant between the two systems 
and its value can be inferred from experimental quantities.

Two capacitively coupled JJs can be described in terms of two interacting qubits, provided that the higher energy levels can be ignored in the dynamic evolution.
Considering both the JJ and the axion as two effective TLS, the analogous spin-like Hamiltonian for the quantized axion-JJ system, in unit of $\hslash$, can be assumed to have the following form
\begin{equation}
H_{Ja} = \omega_J \hat{\sigma}_J^z + \omega_a \hat{\sigma}_a^z + \gamma \hat{\sigma}_J^x \hat{\sigma}_a^x.
\label{axion-JJ Hamiltonian}
\end{equation}
The above equation represents a two spin-1/2 system subjected to two local effective fields ($\omega_J$ and $\omega_a$) and coupled through the simpler Heisenberg interaction of strength $\gamma$.

\subsection{Axion-induced Oscillations}

The dynamical problem related to the above model of Eq.~\eqref{axion-JJ Hamiltonian} can be exactly solved; the energy spectrum and the time evolution operator can be analytically derived.
The symmetry property of the Hamiltonian $H_{Ja}$ allows to reduce the system dynamics to that of two independent TLSs.
In other words, the existence of a constant of motion generates two dynamically invariant two-dimensional subspaces, within which the system can be effectively described in terms of a single 
TLS.
In this way the solution of two effective independent two-level dynamics is required to solve the dynamics of the axion-Josephson system  (see Appendix \ref{App}).

Of course, if the condition $k_BT \ll \omega_J$ is satisfied and the axion-JJ coupling is zero ($\gamma=0$), the JJ would remain in its initial state.
Conversely, if $\gamma \neq 0$, the transition might occur for the presence of the axions.
Therefore, an axion-induced dynamics, in which clear signatures can be detected through local measures on the JJ, can emerge.
Given the possibility of controlling the quantum state of a JJ, the latter can be assumed initially prepared in the state $\rho_J=\projector{-}$, ($\hat{\sigma}^z \ket{\pm} = \pm \ket{\pm}$), while it is 
reasonable to consider the axion in a generic classical mixture
\begin{equation} \label{Mixed state}
\rho_a = p~\projector{+} + (1-p) \projector{-},
\end{equation}
where $p \in [0,1]$ is the probability that the axion is in the state $\projector{+}$.
In this case, the population $\rho_J^{11}(t)=\bra{+}\rho_J(t)\ket{+}$, in the experimentally reasonable limit $\gamma \ll \omega_J,\omega_a$ and for $\omega_J \simeq \omega_a$, varies over the time as (see Appendix \ref{App})
\begin{equation}
\rho_J^{11}(t) = p~\sin^2 \left( \gamma t \right)
\end{equation}
and the `magnetization' of the JJ is thus given by
\begin{equation}
M_J^z(t) = \text{Tr} \{ \rho_J(t)\hat{\sigma}_J^z \} = 2p\sin^2(\gamma t) - 1.
\end{equation}
In the prescribed limits, the oscillations induced on the JJ by the presence of the axion clearly emerge.
Furthermore, the frequency that characterizes these oscillations is, and more in general depends on, the coupling parameter.
For this reason, by analysing the characteristic period of such 
a peculiar, periodically oscillating behaviour, the estimate of $\gamma$ can be deduced. In this way, consequently, through appropriate experimental tests and investigations, based on the presented scheme, it could be possible, in principle, to reveal the presence of the axion dark matter.
Analogous results are obtained for a generic quantum state ($\ket{\psi_a} = \alpha\ket{+} + \beta\ket{-}, \quad |\alpha|^2+|\beta|^2=1$) for the axion (see Appendix \ref{App}).

The condition $\omega_J \simeq \omega_a$ is essential for the oscillations to occur more clearly.
In general, in fact, the expression of $\rho_J^{11}(t)$ is slightly more complicated, since several contributions with different amplitudes and frequencies are present [see Eqs. \eqref{Prob gen case 1}].
Since by hypothesis $\gamma \ll \omega_J,\omega_a$, if $\omega_J \neq \omega_a$ the oscillations are hidden and impossible to observe because all the contributions are practically negligible. In the resonance condition, on the other hand, there is a term that gives a non-zero contribution (see Appendix \ref{App}). 
Therefore, the matching between the axion and the JJ frequencies is the basic condition for the emergence of a detectable axion-induced dynamics on the JJ.
It is interesting to note that the same condition arises also in other contexts \cite{Bec13,Gri22} as the fundamental constraint for the emergence of axion-induced phenomena in Josephson systems.

However, it should be noted that, at finite temperatures, that is in the case of thermal state $p= 1/[1+\exp\{\omega_a/k_BT\}]$, the condition $\omega_a \gg k_B T$ implies a vanishing parameter $p$, and the axion is in its down state $\rho_a = \projector{-}$.
This is reasonable since, for very low temperatures, thermal transitions are forbidden and the axion has a much greater probability of being in its down state.
The condition $\gamma \ll \omega_J,\omega_a$ freezes the axion-JJ system in its initial state, and the measurable observables of the system show no temporal changes.
\begin{figure}[t!!]
\centering
{\includegraphics[width=0.4\textwidth]{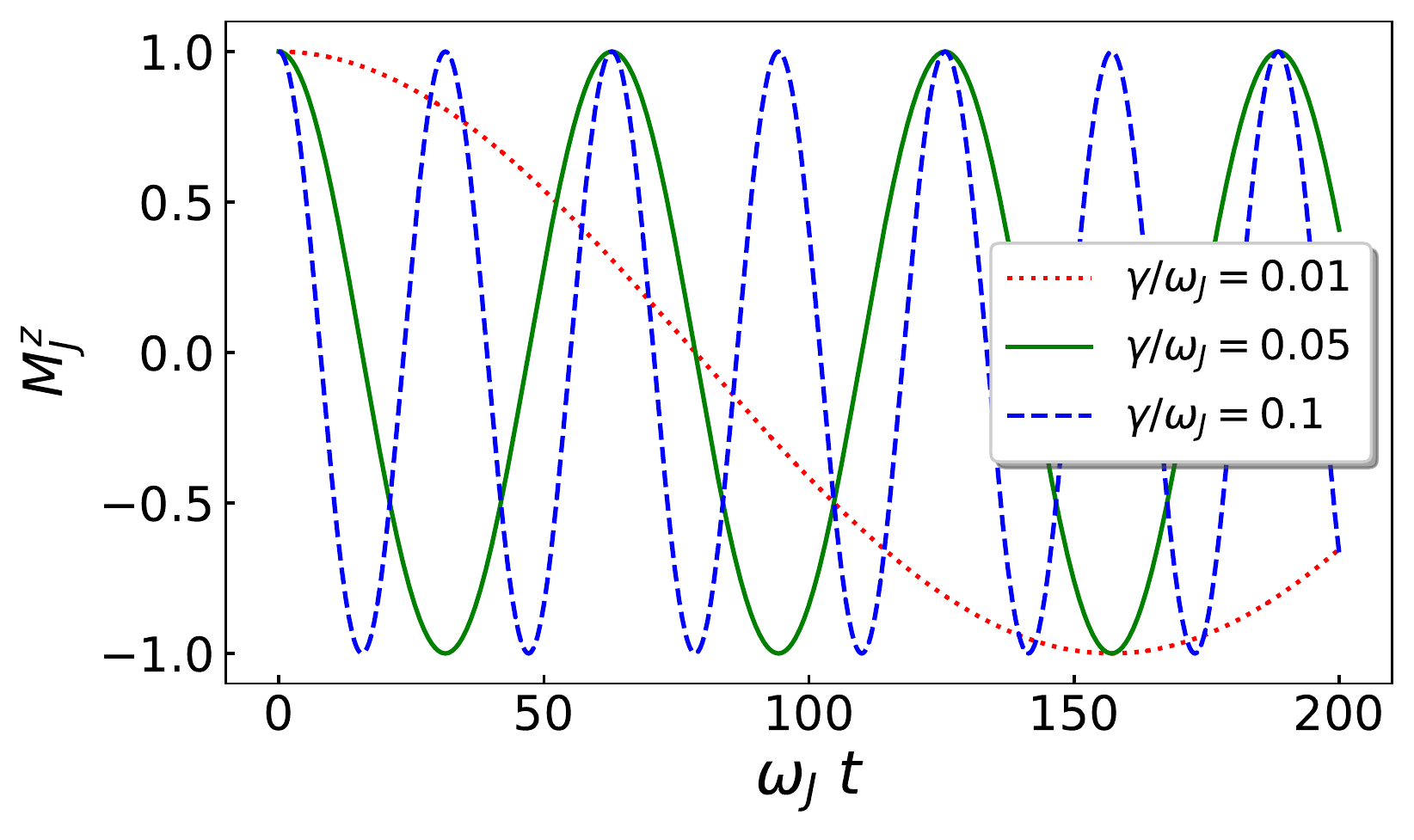}}
\caption{Axion-induced oscillations of the JJ magnetization when the JJ starts from $\rho_J(0)=\ket{+}\bra{+}$ and for three different values of $\gamma/\omega_J$: 0.01 (dotted red line), 0.05 (full green line), 0.1 (dashed blue line).}
\label{fig: ax-jj osc}
\end{figure}
To avoid this problem, it is more convenient to initially prepare the JJ in the up-state $\rho_j(0) = \projector{+}$.
The experimental conditions, $\omega_J \simeq \omega_a$ and $p \approx 0$, imply that (see Appendix \ref{App})
\begin{equation}
\rho_J^{11}(t) = \cos^2 \left( \gamma t \right),
\end{equation}
and consequently the magnetization
\begin{equation}
M_J^z(t) = \cos(2\gamma t),
\end{equation}
presents a periodic reversal phenomenon, shown in Fig.~\ref{fig: ax-jj osc} for three different values of $\gamma$.
Therefore, in addition to the synchronization regime ($\omega_J \simeq \omega_a$), the initial state of the controllable JJ is also essential to enhance the axion-induced oscillations in the junction.

\begin{figure}[b!!]
\centering
{\includegraphics[width=0.4\textwidth]{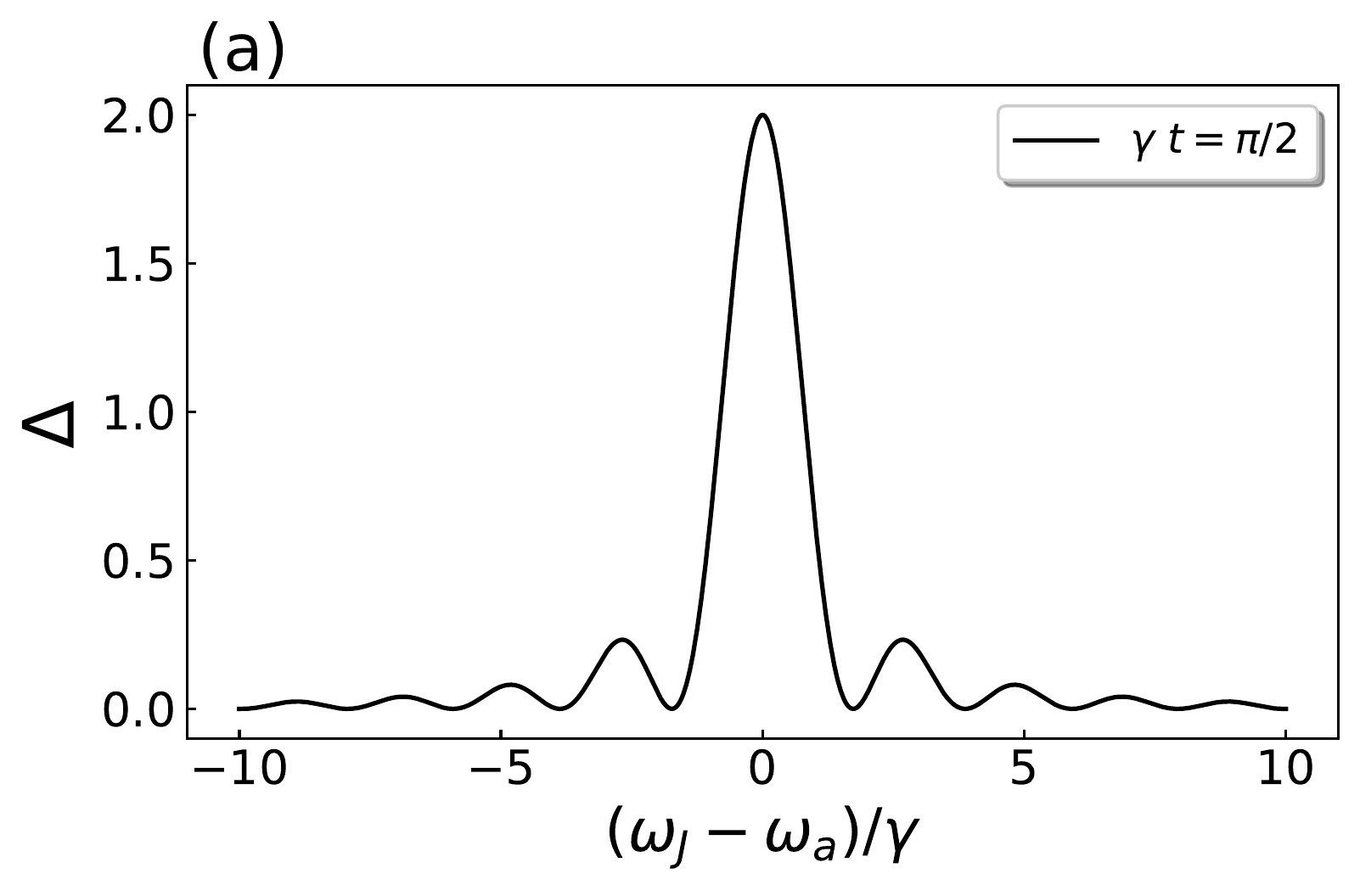}}
\qquad
{\includegraphics[width=0.4\textwidth]{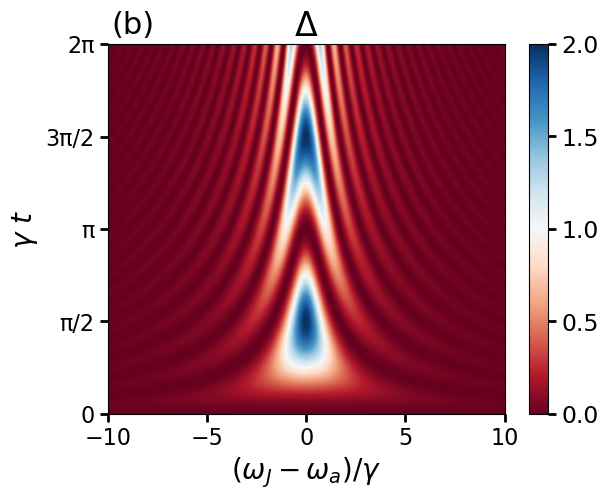}
}
\caption{(a) The amplitude $\Delta \equiv \text{max}(M_J^z)-\text{min}(M_J^z)$ of the axion-induced oscillations of the JJ magnetization as a function of the frequency difference $\omega_J-\omega_a$ for the same initial condition of Fig. \ref{fig: ax-jj osc} [$M_J^z(t=0)=1$], and at the fixed time-instant $\gamma~t = \pi/2$; (b) The same quantity $\Delta$ plotted in the phase space region $\{\gamma~t;~(\omega_J-\omega_a)/\gamma\} = \{[-10,~10];~[0,~2\pi]\}$.
The resonant condition is clearly visible at $\omega_a = \omega_J$ where, for $\gamma~t=(2k+1)\pi/2, ~ k \in \mathrm{N}$, the transition of the JJ from the up- to the down-state occurs more clearly.}
\label{resonance}
\end{figure}

In Fig. \ref{resonance}(a) the amplitude $\Delta \equiv \text{max}(M_J^z)-\text{min}(M_J^z)$ of the axion-induced oscillations of the JJ magnetization at the fixed time-instant $\gamma~t = \pi/2$ as a function of the axion-JJ frequency difference $\omega_J-\omega_a$ is shown.
The same quantity is plotted in Fig. \ref{resonance}(b) in the phase space $\{\gamma~t;~(\omega_J-\omega_a)/\gamma\}$.
The resonant condition due to the frequency matching between the axion and the JJ is clearly visible at $\gamma~t=(2k+1)\pi/2, ~ k \in \mathrm{N}$, where the transition of the JJ from its initial up-state to the down-state occurs.
This resonant dynamical regime reveals a detectable axion-induced dynamics in the JJ, and can be exploited for axion detection.

\section{Conclusions}

In this paper we have considered a JJ coupled to an axion particle. Both systems, at very low temperatures, should dynamically behave as effective TLSs.
The classical model describing the axion-JJ system is analogous to that of two capacitively coupled JJs, which can be quantized, resulting in a two coupled spin-1/2 model.
The close analogy between the axion-JJ system and the two coupled JJs allows to exploit the qubit-qubit Hamiltonian to describe the axion-JJ system as two interacting TLSs.
By solving the dynamic problem exactly, we have highlighted the existence of axion-induced oscillations in the Josephson qubit.
These oscillations can be enhanced by suitably preparing the JJ in its up-state, and matching the Josephson frequency $\omega_J$ with the axion frequency $\omega_a$ (synchronization regime).
Furthermore, the last condition makes the oscillation frequency linearly dependent on the axion-JJ coupling.
Assuming a very low axion-JJ coupling reasonably involves the need for long-term experimental investigations.
Given the typical duration of experiments based on Josephson-qubit~\cite{Blais21, Krantz}, it could be hypothesized that a possible cause of the missing traces of the axion-induced oscillations is the limited stability of the experiments conducted so far.
Therefore, long-term experiments, in the quantum limit of low noise and weak coupling with the environment, can be aimed at identifying these oscillations, and at the same time they could reveal the presence of axion particles and could provide information on the coupling parameter.

\section*{Acknowledgments}
RG acknowledges support by the PRIN Project PRJ-0232 - Impact of Climate Change on the biogeochemistry of Contaminants in the Mediterranean sea (ICCC).
GF acknowledges support by Italian Institute of Nuclear Physics (INFN) through the DARTWARS project.
BS acknowledges support by Government of the Russian Federation through Agreement No. 074-02-2018-330 (2).
All the authors acknowledge the support of the Ministry of University and Research of Italian Government.

\section*{Appendices}


\appendix

\section{Quantized axion-JJ system} \label{App}

Let us consider the following model (in units of $\hslash$)
\begin{equation} 
\begin{aligned}
{H} = 
\omega_{J}\hat{\sigma}_{J}^{z}+\omega_{a}\hat{\sigma}_{a}^{z}+
\gamma \hat{\sigma}_{J}^{x}\hat{\sigma}_{a}^{x}
\label{Hamiltonian}
\end{aligned}
\end{equation}
describing two interacting spin-1/2's subjected to two local longitudinal (along the $z$ direction) fields ($\omega_1$ and $\omega_2$). $\hat{\sigma}_{i}^{x}$ and $\hat{\sigma}_{i}^{z}$ ($i=1,2$) are the Pauli matrices. 

The Hamiltonian exhibits the following canonical symmetry transformation, that is it remains unchanged under such a transformation \cite{GMN, GVM, GIMGM, GNMV},
 \begin{equation}\label{Symmetry Canonical Transformation}
\hat{{\sigma}}_{i}^{x}\to-\hat{\sigma}_{i}^{x},\quad
\hat{{\sigma}}_{i}^{z}\to\hat{\sigma}_{i}^{z},\quad i = J,a.
\end{equation}
This fact implies the existence of a unitary time-independent operator accomplishing the transformation \eqref{Symmetry Canonical Transformation}, which is 
by construction a constant of motion. This unitary operator is given by $\pm \hat{\sigma}_{J}^{z} \hat{\sigma}_{a}^{z}$, being the transformation~\eqref{Symmetry Canonical Transformation} nothing but the rotations of $\pi$ around the $\hat{z}$ axis with respect to each spin. The unitary operator accomplishing this transformation is
\begin{equation}\label{Rotation Operator}
e^{i \pi \hat{S}_{1}^{z} / \hslash} \otimes e^{i \pi \hat{S}_{2}^{z} / \hslash} 
= - \hat{\sigma}_{J}^{z} \hat{\sigma}_{a}^{z} = \cos \Bigl( {\pi \over 2} \hat{\Sigma}_{z} \Bigr),
\end{equation}
where $\hat{\Sigma}_{z} \equiv \hat{\sigma}_{J}^{z} + \sigma_{a}^{z}$. Equation \eqref{Rotation Operator} shows that the constant of motion 
$\hat{\sigma}_{J}^{z} \hat{\sigma}_{a}^{z}$ is indeed a $\hat{\Sigma}_{z}$-based parity operator since, in correspondence of its integer eigenvalues $M = +2,0,0,-2$, it has eigenvalues $1,-1,-1,1$, respectively.

This constant of motion implies two dynamically invariant sub-dynamics related to the two eigenvalues of $\hat{\sigma}_{J}^{z} \hat{\sigma}_{a}^{z}$. We can extract 
these two sub-dynamics by considering that the operator $\hat{\sigma}_{J}^{z} \hat{\sigma}_{a}^{z}$ has the same spectrum of $\hat{\sigma}_{a}^{z}$, i.e., the same 
eigenvalues ($\pm 1$) with the same twofold degeneracy. There exists therefore a unitary time-independent operator $\mathbb{U}$ transforming 
$\hat{\sigma}_{J}^{z} \hat{\sigma}_{a}^{z}$ in $\hat{\sigma}_{a}^{z}$. It can be easily seen that the unitary and Hermitian operator
\begin{equation}
T=\dfrac{1}{2}\left[\mathbb{1}+\hat{\sigma}_{J}^{z}+\hat{\sigma}_{a}^{x}-\hat{\sigma}_{J}^{z}\hat{\sigma}_{a}^{x}\right]
\end{equation}
accomplishes the desired transformation:
\begin{equation}
T^{\dagger}\hat{\sigma}_{J}^{z}\hat{\sigma}_{a}^{z}T=T\hat{\sigma}_{J}^{z}\hat{\sigma}_{a}^{z}T=\hat{\sigma}_{a}^{z}.
\end{equation}
Transforming ${H}$ into ${\tilde{H}} = T^{\dagger} {H} T$, we get
\begin{equation} \label{H tilde}
\begin{aligned}
{\tilde{H}} =
\omega_{J} \hat{\sigma}_{J}^{z} + \omega_{a} \hat{\sigma}_{J}^{z} \hat{\sigma}_{a}^{z} +
\gamma \hat{\sigma}_{J}^{x}.
\end{aligned}
\end{equation}

Therefore $\hat{\sigma}_{a}^{z}$ is a constant of motion of $\tilde{H}$ and, consequently, $\tilde{H}$ may be written parametrically in $\sigma_{a}^{z}$, 
by considering $\hat{\sigma}_{a}^{z}$ as a parameter
\begin{equation}
\begin{aligned}
H_{\sigma_a^z} =
\left(\omega_{J}+\omega_{a}\sigma_{a}^{z}\right)\hat{\sigma}_{J}^{z} + \gamma \hat{\sigma}_{J}^{x}.
\end{aligned}
\end{equation}
This implies the existence of two ($\sigma_{a}^{z} = \pm 1$) two-dimensional sub-dynamics related to the two dynamically invariant Hilbert subspaces.
The two-spin system, thus, in each dynamically invariant subspace behaves effectively as a TLS.

However, it is worth pointing out that the last Hamiltonian is obviously a four-dimensional Hamiltonian since the operators of the first spin are to be understood as multiplied by the identity operator of the second spin ($\sigma_a^z$ is just a number).
Moreover, we may write two two-dimensional Hamiltonians of a fictitious TLS.
In particular, when $\sigma_{a}^{z}=\pm 1$, we get
\begin{equation}\label{eff ham 1}
\begin{aligned}
H_\pm =
\left(\omega_{J} \pm \omega_{a}\right)\hat{\sigma}^{z} + \gamma \hat{\sigma}^{x}.
\end{aligned}
\end{equation}
These two Hamiltonians must be understood as effective two-dimensional Hamiltonians that govern the dynamics of the two-spin system within each dynamically invariant 
two-dimensional Hilbert subspace relative to one of the two eigenvalues of $\hat{\sigma}_a^z$. The overall dynamics of the two-spin system, therefore, in each subspace is 
equivalent to that of a fictitious single spin-1/2 immersed in a fictitious field, which is also coupled to a reservoir through effective coupling constants.

In particular, the subspace related to the eigenvalue $+1$ of  $\hat{\sigma}_{a}^{z}$ is spanned by the two-spin states $\{\ket{++},\ket{--}\}$ ($\hat{\sigma}^z\ket{\pm}=\pm\ket{\pm}$) and the dynamics is ruled by the effective Hamiltonian $H_+$.
It means that the two states $\{\ket{++},\ket{--}\}$ are mapped into the states $\{\ket{+},\ket{-}\}$ 
of the fictitious TLS.
So, in this case, by studying the dynamics of the fictitious TLS, we study the dynamics of the TLS within the space spanned by $\{\ket{++},\ket{--}\}$.
Analogously, the subspace related to the eigenvalue $-1$ of  $\hat{\sigma}_{a}^{z}$ is spanned by the two-spin states $\{\ket{+-},\ket{-+}\}$ and the effective two-level Hamiltonian, ruling the dynamics, is given by $H_-$.
In this case, the two states $\{\ket{+},\ket{-}\}$ of the fictitious spin-1/2 are the mapping images of the two two-spin states $\{\ket{+-},\ket{-+}\}$.

It is important to point out that, even if we describe the dynamics within each subspace in terms of a single fictitious spin-1/2, it does not means that the two actual TLSs are effectively decoupled.
Actually, the dynamics happening within each subspace involves both the real spins; it can be easily seen by the fact that the spin states involved in each dynamically invariant subspace are two-spin states: $\{\ket{++},\ket{--}\}$ in a subspace and $\{\ket{+-},\ket{-+}\}$ in the other one.

Let us consider the two two-dimensional Hamiltonians
\begin{equation}
H_\pm=%
\begin{pmatrix}
\Omega_\pm & \gamma\\
\gamma & -\Omega_\pm
\end{pmatrix}
,\label{SU(2) Hamiltonian}
\end{equation}
where $\Omega_\pm=\omega_J \pm \omega_a$.
Each Hamiltonian represents a spin-1/2 or more in general a TLS subjected to two time-dependent magnetic fields: one along the $z$-direction [giving rise to the energy term $\hslash\Omega(t)$], the other along the $x$-direction [giving rise to the energy term $\hslash\gamma(t)$].
It is important to point out that in the two-level representation what we call transverse field (along the $x$-direction) represents the coupling between the two real interacting spins.
The solution of each (sub)dynamical problem is related to the solution of the Schr\"odinger equation $i\dot{U}_\pm=H_\pm U_\pm$ with $U_\pm(0)=1$.
Here, $U_\pm$ are the time evolution operators generated by $H_\pm$, and can be represented as
\begin{equation}
U_\pm(t)=
\begin{pmatrix}
a_\pm(t) & b_\pm(t)\\
-b_\pm^{\ast}(t) & a_\pm^{\ast}(t)
\end{pmatrix}
,\quad|a_\pm|^{2}(t)+|b_\pm|^{2}(t)=1,\label{Time Ev Op SU2}
\end{equation}
where $a_\pm(t)$ and $b_\pm(t)$ are two complex-valued functions whose expressions can be determined by solving the Schr\"odinger equation. The time evolution operator 
related to the two-spin Hamiltonian, represented in the coupled basis, takes the form
\begin{equation}
U(t)=
\begin{pmatrix}
a_+(t) & 0 & 0 & b_+(t) \\ 
0 & a_-(t) & b_-(t) & 0 \\
0 & -b_-(t) & a_-^*(t) & 0 \\
-b_+^*(t) & 0 & 0 & a_+^*(t)
\end{pmatrix}.
\end{equation}
It is possible to easily get the two exact expressions of $a_\pm(t)$ and $b_\pm(t)$ which read
\begin{subequations}\label{No time Solutions}
\begin{align}
a_\pm(t)&= \left[ \cos \left( \nu_\pm t \right)-i{\Omega_\pm \over \nu_\pm}\sin \left( \nu_\pm t \right) \right], \\
b_\pm(t)&=-i{\gamma \over \nu_\pm} \sin \left( \nu_\pm t \right),
\end{align}
\end{subequations}
with $\nu_\pm=\sqrt{\Omega_\pm^2+\gamma^2}$.

If the two real spins are initialized in $\ket{--}$, that is
\begin{equation}
\rho_{Ja}(0)=
\begin{pmatrix}
0 & 0 & 0 & 0 \\ 
0 & 0 & 0 & 0 \\
0 & 0 & 0 & 0 \\
0 & 0 & 0 & 1
\end{pmatrix},
\end{equation}
the dynamics of the two-spin system is restricted to the two-dimensional subspace spanned by $\{ \ket{++},\ket{--} \}$ and then only the solution of the dynamical 
problem related to $H_+$ is required. In this case, the initial state of the effective TLS, according to the underlying mapping, is $\rho_+(0)=\ket{-}\bra{-}$.
Therefore, the entries of the matrix
\begin{equation}
\rho_+(t) = U_+(t) \rho_+(0) U_+^\dagger(t) =
\begin{pmatrix}
\rho_{11}^+(t) & \rho_{12}^+(t) \\
\rho_{21}^+(t) & \rho_{22}^+(t)
\end{pmatrix}
\end{equation}
are exactly the four entries of the two-(real)spin matrix
\begin{equation}\label{rho_s(t)}
\rho_{Ja}(t) = U(t) \rho_{Ja}(0) U^\dagger(t) =
\begin{pmatrix}
\rho_{JJ}^+(t) & 0 & 0 & \rho_{Ja}^+(t) \\ 
0 & 0 & 0 & 0 \\
0 & 0 & 0 & 0 \\
\rho_{aJ}^+(t) & 0 & 0 & \rho_{aa}^+(t)
\end{pmatrix}.
\end{equation}
Of course, the analogous result is obtained if the two spins start from $\ket{-+}$. This implies that only the subdynamics governed by $H_-$ is involved and then, 
on the basis of the mapping, we get the following evolved density matrix
\begin{equation}\label{rho_s(t) new}
\rho_{Ja}(t) = U(t) \rho_{Ja}(0) U^\dagger(t) =
\begin{pmatrix}
0 & 0 & 0 & 0 \\ 
0 & \rho_{JJ}^-(t) & \rho_{Ja}^-(t) & 0 \\
0 & \rho_{aJ}^-(t) & \rho_{aa}^-(t) & 0 \\
0 & 0 & 0 & 0
\end{pmatrix},
\end{equation}
where the entries are those of the effective two-dimensional state $\rho_-(t) = U_-(t) \rho_-(0) U_-^\dagger(t)$.

Therefore, if the two spins start from the state
\begin{equation}
\rho_{Ja}(0) = 
\begin{pmatrix}
0 & 0 & 0 & 0 \\ 
0 & 0 & 0 & 0 \\
0 & 0 & |\alpha|^2 & \alpha\beta^* \\
0 & 0 & \alpha^*\beta & |\beta|^2
\end{pmatrix}
= \rho_J(0) \otimes \rho_a(0)
\end{equation}
with
\begin{eqnarray}
\rho_J(0) &=&
\begin{pmatrix}
0 & 0  \\
0 & 1
\end{pmatrix}
= \ket{-}\bra{-},  \nonumber \\
\rho_a(0) &=&
\begin{pmatrix}
|\alpha|^2 & \alpha\beta^* \\
\alpha^*\beta & |\beta|^2
\end{pmatrix}
= \ket{\psi_a}\bra{\psi_a}, \quad \ket{\psi_a} = \alpha\ket{+} + \beta\ket{-}, \nonumber \\
&& |\alpha|^2+|\beta|^2=1,
\end{eqnarray}
one gets
\begin{eqnarray} \label{rho11 1st case}
\rho_J(t) &=& \text{Tr}_a\{ \rho_{Ja}(t) \}  \nonumber \\
&=& (|b_+|^2 |\beta|^2 + |b_-|^2 |\alpha|^2) \projector{+} + \nonumber \\
&& - (b_+ a_-^* \alpha^* \beta + a_+^* b_- \alpha \beta^*) \ketbra{+}{-} + \nonumber \\
&& - (b_+^* a_- \alpha \beta^* + a_+ b_-^* \alpha^* \beta) \ketbra{-}{+} +  \nonumber \\
&& + (|a_+|^2 |\beta|^2 + |a_-|^2 |\alpha|^2) \projector{-}.
\end{eqnarray}

If, on the other hand, the initial state $\rho_{Ja}(0) = \rho_J(0) \otimes \rho_a(0)$ of the two spins is defined by
\begin{eqnarray}
\rho_J(0) &=&
\begin{pmatrix}
0 & 0 \\
0 & 1
\end{pmatrix}
= \ket{-}\bra{-}, \nonumber \\
\rho_a(0) &=&
\begin{pmatrix}
p & 0 \\
0 & 1-p
\end{pmatrix}
= p~\projector{+} + (1-p) \projector{-}, \nonumber \\
&& p \in [0,1]
\end{eqnarray}
the reduced density matrix of the first spin becomes
\begin{eqnarray} \label{rho11 2nd case}
\rho_J(t) &=& \text{Tr}_a\{ \rho_{Ja}(t) \} \nonumber \\
&=& [(1-p)|b_+|^2 + p |b_-|^2] \projector{+} + \nonumber \\
&& + [(1-p)|a_+|^2 + p |a_-|^2 ] \projector{-}.
\end{eqnarray}
The probability $\bra{+}\rho_J(t)\ket{+} \equiv \rho_J^{11}(t)$ of the transition $\ket{-} \rightarrow \ket{+}$ for the JJ thus reads
\begin{subequations} \label{Prob gen case 1}
  \begin{align}
    \rho_J^{11}(t) &= (1-p)|b_+|^2 + p |b_-|^2, \\
    |b_\pm|^2 &= {\gamma^2 \over \nu_\pm^2} \sin^2(\nu_\pm~t), \\
    \nu_\pm &= \sqrt{\gamma^2 + \Omega_\pm^2}.
  \end{align}    
\end{subequations}
We see that, in the limit $\gamma \ll \omega_J,\omega_a$, and for $\omega_J \simeq \omega_a$, we have that $\gamma/\nu_+ \approx 0$ and $\gamma/\nu_- \approx 1$. 
So that $b_+ \approx 0$ and the transition probability takes the simple form
\begin{equation}
\rho_J^{11}(t) = |\alpha|^2~\sin^2 \left( \gamma t \right)
\quad \text{and} \quad
\rho_J^{11}(t) = p~\sin^2 \left( \gamma t \right)
\end{equation}
in the first [Eq. \eqref{rho11 1st case}] and the second [Eq. \eqref{rho11 2nd case}] case, respectively.
In the second case, if $\rho_J(0) = \projector{+}$, we obtain
\begin{eqnarray}
\rho_J(t) &=& \text{Tr}_a\{ \rho_{Ja}(t) \} \nonumber \\
&=& [p|a_+|^2 + (1-p) |a_-|^2 ] \projector{+} + \nonumber \\
&& + [p|b_+|^2 + (1-p) |b_-|^2]\projector{-}.
\end{eqnarray}
Thus, for $p \approx 0$, $\gamma \ll \omega_J,\omega_a$, and $\omega_J \simeq \omega_a$, we obtain [$\rho_J^{22}(t) \equiv \bra{-}\rho_1(t)\ket{-}$]
\begin{subequations}
  \begin{align}
    \rho_J^{11}(t) &= \cos^2(\gamma t),
    \\
    \rho_J^{22}(t) &= \sin^2(\gamma t),
    \\
    M_J^z(t) &= \rho_J^{11}(t) - \rho_J^{22}(t) = \cos(2 \gamma t).
  \end{align}
\end{subequations}

Finally, it is worth pointing out that the symmetry property possessed by the Hamiltonian does not depend on the Hamiltonian parameters, but only on its structure.
This implies that even in the presence of time-dependent characteristic frequencies of the spins [$\omega_J(t)$ and $\omega_a(t)$], the reduction of the main two-spin dynamical problem to two independent effective single two-level dynamical problems remain valid. 


%

\end{document}